\documentclass[twocolumn,twocolappendix]{aastex63}

\received{}
\revised{}
\accepted{}
\shorttitle{X-ray Observations of J0439+1634}
\shortauthors{Yang et al.}

\def\mgii{\ion{Mg}{2}}
\def\civ{\ion{C}{4}}

\usepackage{threeparttable}

\begin{document}

\title{Deep {\em XMM-Newton} Observations of an X-ray Weak, Broad Absorption Line Quasar at $z=6.5$}

\correspondingauthor{Jinyi Yang}
\email{jinyiyang@email.arizona.edu}

\author[0000-0001-5287-4242]{Jinyi Yang}
\altaffiliation{Strittmatter Fellow}
\affiliation{Steward Observatory, University of Arizona, 933 N Cherry Avenue, Tucson, AZ 85721, USA}

\author[0000-0003-3310-0131]{Xiaohui Fan}
\affiliation{Steward Observatory, University of Arizona, 933 N Cherry Avenue, Tucson, AZ 85721, USA}

\author[0000-0002-7633-431X]{Feige Wang}
\altaffiliation{NASA Hubble Fellow}
\affiliation{Steward Observatory, University of Arizona, 933 N Cherry Avenue, Tucson, AZ 85721, USA}

\author[0000-0001-9094-0984]{Giorgio Lanzuisi}
\affiliation{INAF -- Osservatorio di Astrofisica e Scienza dello Spazio di Bologna, via Gobetti 93/3, I-40129 Bologna, Italy}

\author[0000-0002-2579-4789]{Riccardo Nanni}
\affiliation{Leiden Observatory, Leiden University, P.O. Box 9513, NL-2300 RA Leiden, The Netherlands}

\author[0000-0001-6966-8920]{Massimo Cappi}
\affiliation{INAF -- Osservatorio di Astrofisica e Scienza dello Spazio di Bologna, via Gobetti 93/3, I-40129 Bologna, Italy}

\author[0000-0003-1697-6596]{George Chartas}
\affiliation{Department of Physics and Astronomy of the College of Charleston, Charleston, SC 29424, USA}

\author[0000-0002-7858-7564]{Mauro Dadina}
\affiliation{INAF -- Osservatorio di Astrofisica e Scienza dello Spazio di Bologna, via Gobetti 93/3, I-40129 Bologna, Italy}

\author[0000-0002-2662-8803]{Roberto Decarli}
\affiliation{INAF -- Osservatorio di Astrofisica e Scienza dello Spazio di Bologna, via Gobetti 93/3, I-40129 Bologna, Italy}

\author[0000-0002-5768-738X]{Xiangyu Jin}
\affiliation{Steward Observatory, University of Arizona, 933 N Cherry Avenue, Tucson, AZ 85721, USA}

\author[0000-0001-6812-2467]{Charles R. Keeton}
\affiliation{Department of Physics and Astronomy, Rutgers University, Piscataway, NJ 08854, USA}

\author[0000-0001-9024-8322]{Bram P.\ Venemans}
\affiliation{Max-Planck Institute for Astronomy, K{\"o}nigstuhl 17, D-69117 Heidelberg, Germany}

\author[0000-0003-4793-7880]{Fabian Walter}
\affiliation{Max-Planck Institute for Astronomy, K{\"o}nigstuhl 17, D-69117 Heidelberg, Germany}

\author[0000-0003-4956-5742]{Ran Wang}
\affiliation{Kavli Institute for Astronomy and Astrophysics, Peking University, Beijing 100871, China}

\author[0000-0002-7350-6913]{Xue-Bing Wu}
\affiliation{Kavli Institute for Astronomy and Astrophysics, Peking University, Beijing 100871, China}

\author[0000-0002-5367-8021]{Minghao Yue}
\affiliation{Steward Observatory, University of Arizona, 933 N Cherry Avenue, Tucson, AZ 85721, USA}

\author[0000-0001-6047-8469]{Ann Zabludoff}
\affiliation{Steward Observatory, University of Arizona, 933 N Cherry Avenue, Tucson, AZ 85721, USA}




\begin{abstract}
We report X-ray observations of the most distant known gravitationally lensed quasar, J0439+1634 at $z=6.52$, which is also a broad absorption line (BAL) quasar, using the {\em XMM-Newton} Observatory. With a 130 ks exposure, the quasar is significantly detected as a point source at the optical position with a total of 358$^{+19}_{-19}$ net counts using the EPIC instrument. By fitting a power-law plus Galactic absorption model to the observed spectra, we obtain a spectral slope of $\Gamma=1.45^{+0.10}_{-0.09}$. The derived optical-to-X-ray spectral slope $\alpha_{\rm{ox}}$ is $-2.07^{+0.01}_{-0.01}$, suggesting that the X-ray emission of J0439+1634 is weaker by a factor of 18 than the expectation based on its 2500 \AA\ luminosity and the average $\alpha_{\rm{ox}}$ vs. luminosity relationship. This is the first time that an X-ray weak BAL quasar at $z>6$ has been observed spectroscopically. Its X-ray weakness is consistent with the properties of BAL quasars at lower redshift.
By fitting a model including an intrinsic absorption component, we obtain intrinsic column densities of $N_{\rm{H}}=2.8^{+0.7}_{-0.6}\times10^{23}\,\rm{cm}^{-2}$ and $N_{\rm{H}}= 4.3^{+1.8}_{-1.5}\times10^{23}\,\rm{cm}^{-2}$, assuming a fixed $\Gamma$ of 1.9 and a free $\Gamma$, respectively. The intrinsic rest-frame 2--10 keV luminosity is derived as $(9.4-15.1)\times10^{43}\,\rm{erg\,s}^{-1}$, after correcting for lensing magnification ($\mu=51.3$). The absorbed power-law model fitting indicates that J0439+1634 is the highest redshift obscured quasar with a direct measurement of the absorbing column density.
The intrinsic high column density absorption can reduce the X-ray luminosity by a factor of $3-7$, which also indicates that this quasar could be a candidate of intrinsically X-ray weak quasar. 
\end{abstract}



\section{Introduction} \label{sec:intro}
Reionization-era quasars are direct probes of super-massive black hole (SMBH) and massive galaxy assembly in the early Universe. Recent successful high-redshift quasar surveys provide a large sample of new reionization-era quasars \citep[e.g.,][]{banados18,reed19,matsuoka19,venemans15,wang19,yang19b}. Near-infrared (NIR) spectroscopy of these quasars has revealed the existence of billion solar mass BHs within the first Gyr after the Big Bang and has also suggested that these SMBHs are accreting close to the Eddington limit \citep[e.g.,][]{shen19,schindler20,yang21}. 
X-ray emission from high-redshift quasars probes the conditions in the innermost regions of their accretion-disk corona, and thus provides information about how these early SMBHs are fed.

There have been substantial efforts to conduct multi-wavelength observations of the most distant quasars; their properties in the optical, NIR to (sub)millimeter bands have been well characterized. However, studies of the X-ray properties of these systems are still limited. There are $\sim$ 30 quasars detected in X-rays at $z>6$ \citep[e.g.,][]{shemmer06,page14,moretti14,ai17,gallerani17,nanni17,nanni18,banados18,vito19a,connor19,connor20,wang21} among more than 200 quasars known at this redshift,  and only seven at $z>6.5$ with X-ray detections \citep[e.g.,][]{page14,vito19a,banados18,connor20,pons20,wang21}.
Due to their extreme distances, the observed X-ray emission of these quasars is very faint, and only four $z>6$ quasars are detected with more than 100 net counts \citep{page14, ai17, nanni17, vito19a}, which limits detailed investigations of their X-ray spectral
properties and any correlations with other quasar properties. 
Recent studies of high-redshift quasars have mostly focused on their average X-ray properties. 
In addition, all previous X-ray observations have been obtained only for the most intrinsically luminous quasars, probing only the most massive SMBHs. 

Quasar J0439+1634 at $z=6.5188$ is the first known gravitationally lensed quasar at $z>5$ and the brightest quasar known at this redshift at rest-frame UV/optical to far-infrared wavelengths due to the lensing. The high lensing magnification \citep[$\mu$=51.3,][]{fan19} makes J0439+1634 an ideal target for the study of X-ray emission from a reionization-era quasar that is intrinsically less luminous. In addition, this quasar is also a broad absorption line (BAL) quasar \citep{yang21}. BAL quasars have been suggested to be
highly absorbed in the soft X-ray band and are generally X-ray weak in observations of low-redshift quasars; no such studies of high-redshift BAL quasars have been carried out, due to their faint X-ray emission. 
In this work, we report {\em XMM-Newton} observations of J0439+1634. We investigate its X-ray properties through spectral analysis and compare it with other high-redshift and low-redshift quasar populations.  
All results below refer to a $\Lambda$CDM cosmology with parameters $\Omega_{\Lambda}$ = 0.7, $\Omega_{m}$ = 0.3, and $h$ = 0.7. 

\section{{\em XMM-Newton} Observations and Data Reduction} \label{sec:obs}
\begin{figure*}%
\centering 
\epsscale{1.2}
\plotone{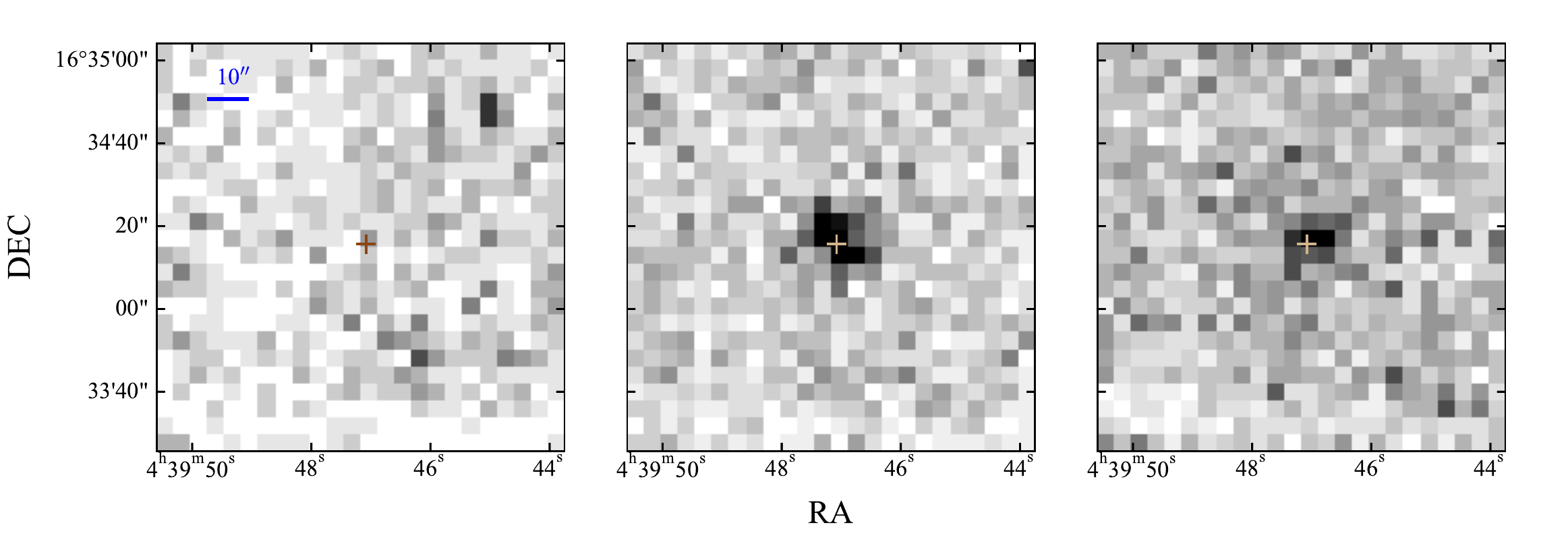} 
\caption{The {\em XMM-Newton} EPIC image ($100'' \times 100''$) of J0439+1634, combined from the pn, MOS1, and MOS2 images, in the 0.2--0.5 keV (left), 0.5--2 keV (middle) and 2--10 keV (right) X-ray bands. The central crosses represent the optical coordinates of this quasar (i.e., J043947.08+163415.7, \citealt{fan19}). The quasar is detected in the 0.5--2 keV (rest-frame 3.8--15 keV) and 2--10 keV (rest-frame 15--75 keV) bands, but not in the 0.2--0.5 keV band (rest-frame 1.5--3.8 keV).}
\label{fig:imaging}
\end{figure*}

J0439+1634 was observed with {XMM-Newton} on 2020 August 24 for a total observing time of 130 ks (Program ID 86320). The European Photon Imaging Camera (EPIC) was operated in full-frame mode, with thin
filters. The EPIC data have been processed with {\em XMM-Newton} Science Analysis Software (SAS) v19.1.0 following the standard data analysis threads \footnote{\url{https://www.cosmos.esa.int/web/xmm-newton/sas-threads}}.
We filtered out time periods with high particle backgrounds by limiting the count-rate thresholds to $<0.4$ cts s$^{-1}$ in the 10 $< E < 12$ keV band light curves for the pn camera and $<0.35$ cts s$^{-1}$ in the $E > 10$ keV band for the MOS cameras.  
We only considered events with patterns 0--4 and 0--12 for the scientific analysis for the pn and MOS cameras, respectively.
We created images and extracted spectra, response matrices, and ancillary files using the {\it evselect}, {\it backscale}, {\it rmfgen}, and {\it arfgen} tools.
For spectral extraction, we extracted the counts from the three cameras separately. We  extracted counts from circular regions centered at the optical position of the quasar with a radius of 12$''$, and we selected an object-free nearby circular regions with a 60$''$ radius for the background, twenty-five times larger than the target extraction area. 
We also group the spectra using {\it specgroup} with a minimum number of counts of one per bin, which will be used for spectral fitting. 
The total effective exposure times are 76.9 ks, 118.7 ks, and 118.6 ks for pn, MOS1, and MOS2, respectively. 
We merged the images from the three EPIC cameras using the {\it merge} tool, in three different observed bands, from soft to hard: 0.2--0.5 keV (rest-frame 1.5--3.8 keV), 0.5--2 keV (rest-frame 3.8--15 keV), and 2--10 keV (rest-frame 15--75 keV), as shown in Figure \ref{fig:imaging}.

\section{Results} \label{sec:results}
\subsection{X-ray Detection and Spectra}
An X-ray source is clearly detected at a position consistent with the optical coordinates of J0439+1634 (i.e., J043947.08+163415.7, \citealt{fan19}) in 0.5--10 keV in all three EPIC cameras, while it is not detected in the 0.2--0.5 keV band (Figure \ref{fig:imaging}) using {\it edetect\_chain}. 
The X-ray source is located 0.7\arcsec\ -- 0.9\arcsec\ away from the optical coordinates in the pn and MOS1 images and 1.5\arcsec\ away in MOS2 (with a $1\sigma$ uncertainty of $\sim 0.8''$ in all three images), based on {\it edetect\_chain}. 
The X-ray and optical coordinates overlap within the spatial resolution in the all three cameras.
The target is detected with false source probabilities of less than $3\times10^{-8}$ by all three cameras, derived using the binomial no-source probability \citep{weisskopf07,broos07,vito19a}.
In this lensing system, the foreground lensing galaxy is 0$\farcs$5 to the East from the quasar, and the separation between the two lensed quasar images is about 0$\farcs$2 in the {\em HST} image \citep{fan19}. Thus, the entire system is unresolved in the EPIC observations. {\em HST} and ground-based observations show that the foreground lens galaxy is a faint, low-mass galaxy (dynamical mass $\sim 2 \times 10^{10}\, M_\odot$ from the best-fit lensing model) without any hints of AGN activity \citep{fan19,yue21}. 
We therefore consider that all X-ray emission detected here is from the quasar.
The number of total net counts from the three cameras is 358$^{+19}_{-19}$ in the 0.2--10 keV band. 
There are no other source detected within 50\arcsec\ of the quasar. 

\begin{figure}%
\centering 
\epsscale{1.21}
\plotone{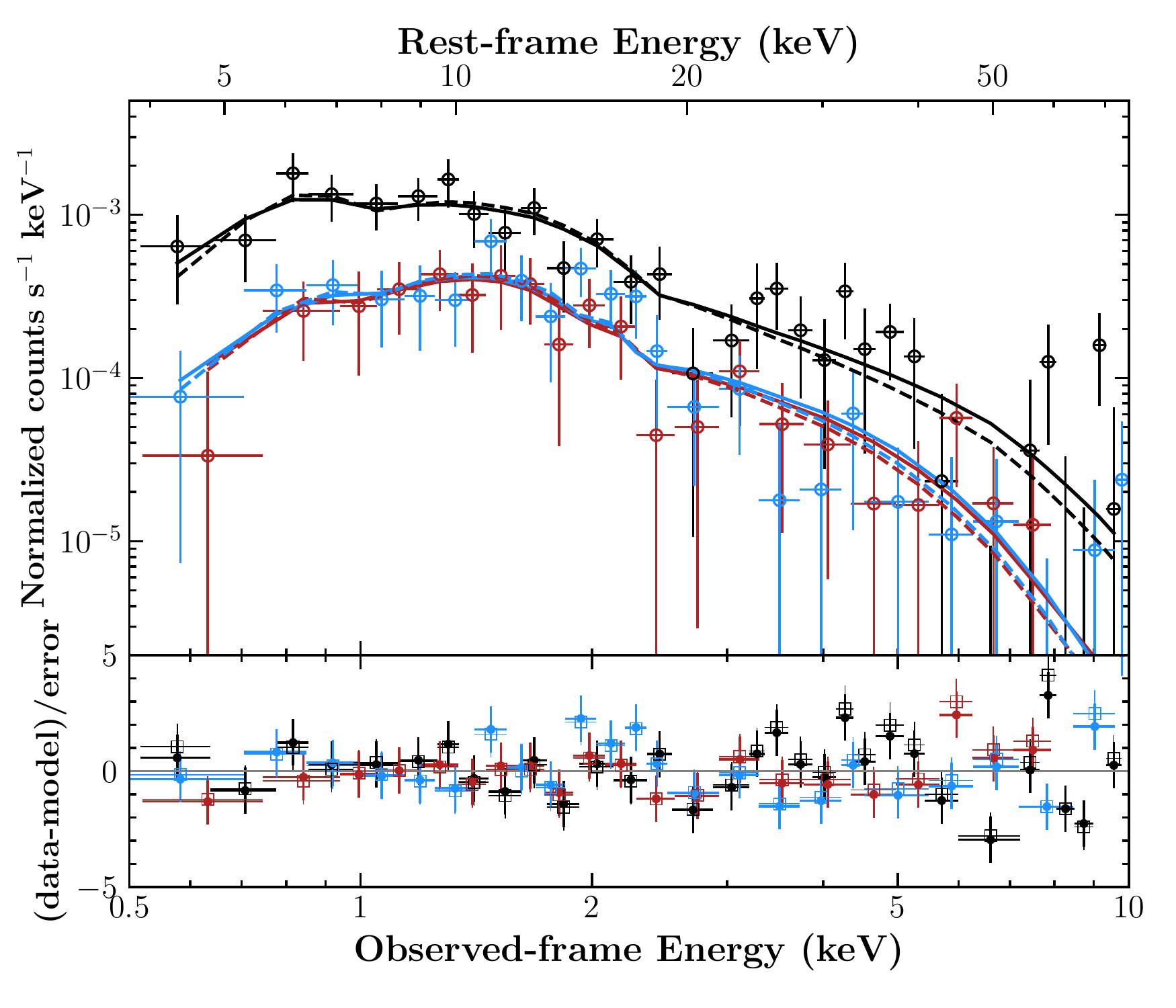} 
\caption{The {\em XMM-Newton} EPIC spectra of quasar J0439+1634. The upper panel shows the observed counts from the pn camera (black), MOS1 (red), and MOS2 (blue). The solid lines represent the spectral fits using an absorbed power-law model consistent of a power-law model, Galactic absorption, and quasar intrinsic absorption, with the assumption of an intrinsic photon index of $\Gamma$ = 1.9 \citep{nanni17}. The dashed lines denote the spectral fits with a free $\Gamma$ (see details of spectral fitting in Section 3.2). The bottom panel shows the residuals, with filled circles for the fixed-$\Gamma$ fits and open squares for the $\Gamma$-free fits. The spectral fitting is applied to the spectra grouped to one count per bin, and the data have been re-binned for plotting purpose here using XSPEC. }
\label{fig:fitting}
\end{figure}

Figure \ref{fig:fitting} shows the spectra from the three EPIC cameras. There is a drop in the counts from $\sim 0.8$ keV to the soft end, which does not appear in other EPIC spectra of $z> 6$ quasars \citep{page14,moretti14,ai17,nanni17} and potentially indicates strong absorption in the rest-frame soft band.
We apply spectral fitting to the data that have been grouped into one count per bin using XSPEC with the Cash statistic \citep{cash79}. 
When fitting, we fix the Galactic HI column density to $N_{\rm H}$ of $1.46 \times 10^{21}$ cm$^{-2}$, derived based on the Leiden/Argentine/Bonn map \citep{kalberla05}. 

We first fit the data with a power-law plus Galactic absorption model ({\tt phabs*zpowerlw} in XSPEC) and obtain a best-fit photon index of $\Gamma$ = 1.45 $^{+0.10}_{-0.09}$, with a C-Statistic of 247.51 for 240 degrees of freedom.
The slope is flatter than the typical slopes for $z \gtrsim 6$ quasars that have X-ray spectra (e.g., $\Gamma \sim 1.6 - 2.2$, \citealt{page14,ai17,nanni17,vito19a}); it is also flatter than the measurements for low-redshift quasars \citep[e.g.,][]{vignali05, just07}. Comparisons between radio-loud and radio-quiet quasars at lower redshift suggest a flatter slope ($\Gamma \sim 1.55$, e.g., \citealt{page05}) for radio-loud quasars than for  radio-quiet quasars ($\Gamma \sim 1.98$, e.g., \citealt{page05}, also \citealt{scott11}). J0439+1634, however, is radio-quiet \citep{yang19a}.

Based on the spectral fitting results with the power-law model, we compute the Galactic absorption-corrected fluxes. The flux in the 0.5--2 keV band is $4.9^{+0.3}_{-0.3} \times 10^{-15}$ $\rm erg\,s^{-1} cm^{-2}$, and the flux in the 2--10 keV band is $13.1^{+0.8}_{-0.8} \times 10^{-15}$ $\rm erg\,s^{-1} cm^{-2}$.
The rest-frame 2--10 keV band luminosity based by our fit is $2.2^{+0.1}_{-0.1} \times 10^{45}$ $\rm erg\,s^{-1}$ without correction for magnification. After correction for magnification ($\mu$=51.3, \citealt{fan19}), the intrinsic rest-frame 2--10 keV band luminosity is $4.3^{+0.3}_{-0.3} \times 10^{43}$ $\rm erg\,s^{-1}$.

We estimate the optical-X-ray power-law slope, $\alpha_{\rm ox}$, based on the flux densities at 2500 \AA\ and 2 keV in the rest frame. The rest-frame 2 keV (0.266 keV in the observed-frame) flux density is calculated using the PIMMS based on the Galactic absorption-corrected flux. The rest-frame 2500 \AA\ flux density is derived from the NIR spectral fitting in \cite{yang21}. We find a slope $\alpha_{\rm ox}$ of $-2.07^{+0.01}_{-0.01}$. We then compare the measurement of J0439+1634 with the existing relations between $\alpha_{\rm ox}$ and 2500 \AA\ luminosity density and measurements in the literature. J0439+1634, with $L_{\rm 2500 \AA} = 8.0 \times 10^{30}\, \rm erg\,s^{-1} Hz^{-1}$ after correcting for the lensing magnification, has an intrinsic luminosity significantly lower than most $z\gtrsim 6$ quasars previously studied in X-rays.

\begin{figure}%
\centering 
\epsscale{1.22}
\plotone{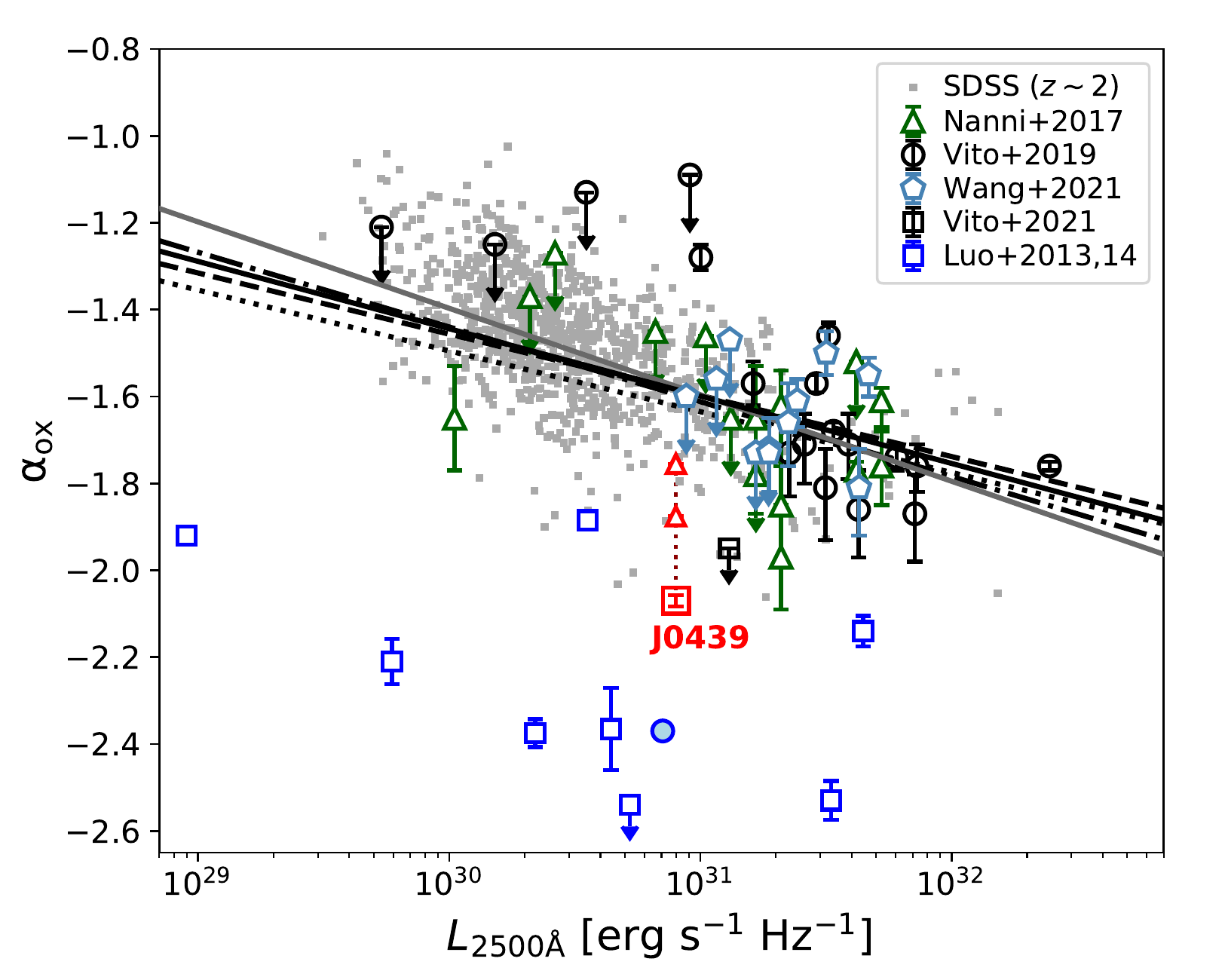} 
\caption{Optical-to-X-ray spectral slope $\alpha_{\rm ox}$ vs. 2500 \AA\ luminosity. Our measurement from J0439+1634 is shown in red, compared with values from quasar samples at similar redhsift and lower redshift. The two red triangles are the $\alpha_{\rm ox}$ values accounting for the intrinsic absorption based on $\Gamma$-fixed (lower value) and $\Gamma$-free (upper value) model fitting. For high-redshift comparisons, we show individual measurements at $5.5 < z < 6$ from \cite{nanni17}, $6 < z < 6.5$ from \cite{vito19a}, and $z> 6.5$ from \cite{wang21}. We also include the new upper limit (black square) from a $z=6.515$ quasar in \cite{vito21}. The low-redshift SDSS quasar sample (grey dots) is generated based on the $L_{\rm 2-10 keV}$ from \cite{timlin20} and $L_{\rm 2500 \AA}$ from \cite{wang21}. The relations in the literature are plotted as black solid line \citep{nanni17}, gray solid line \citep{timlin20}, dashed line \citep{lusso17}, dotted line \citep{just07}, and dash-dot line \citep{martocchia17}. The blue squares are the measurements for low-redshift BAL quasars from \cite{luo14}. Among the sample in \cite{luo14}, the quasar PG0043+03 is not detected, and it has been updated in \cite{kollatschny16} (blue filled circle). All measurements have been corrected to the cosmology adopted in this work.}
\label{fig:aox}
\end{figure}

J0439+1634 is far below the $\alpha_{\rm ox}$--$L_{\rm 2500 \AA}$ relations derived from low-redshift and $z\gtrsim6$ quasar samples in the literature \citep[e.g.,][]{just07,lusso17,martocchia17,nanni17,timlin20,bisogni21}, as shown in Figure \ref{fig:aox}, suggesting a much fainter X-ray luminosity from this quasar relative to its UV luminosity. 
Using the intrinsic continuum luminosity $L_{\rm 2500 \AA}$ of J0439+1634 and applying the existing relation between $\alpha_{\rm ox}$ and 2500 \AA\ luminosity density from \cite{nanni17}, we find that the expected rest-frame 2 keV luminosity should be 18 times higher than the observation. 
J0439+1634 is one of the highest redshift X-ray weak quasars and the only such object with high quality X-ray spectroscopy at $z\gtrsim 6.5$, during the epoch of reionization. 
There are some other high-redshift quasars without X-ray detections, which could thus be candidates of X-ray weak quasars. In particular, \cite{vito21} report an upper limit on X-ray emission from a $z=6.515$ quasar and suggest that its X-ray emission is $> 6$ times weaker than the expectation based on UV luminosity.

\subsection{High Obscuration}
\begin{table*}
      \caption{Best-fit results of EPIC Spectra}
        \centering
         \begin{tabular}{c c c c c c c}
            \hline
            \hline \rule[0.7mm]{0mm}{3.5mm}
            Model & C-stat/d.o.f. & $\Gamma$ & $N\rm_{H}$ & f$_{\rm (0.5-2 \: keV)}$ & f$_{\rm (2-10 \: keV)}$ & L$_{\rm (2-10 \: keV, rest)}$\\
           (1) & (2) & (3) & (4) & (5) & (6) & (7)\\
            \hline \rule[0.7mm]{0mm}{3.5mm}
           Power-law & 247.51/240 & $1.45^{+0.10}_{-0.09}$  & --- & $4.9^{+0.3}_{-0.3}$ & $13.1^{+0.8}_{-0.8}$ & $2.2^{+0.1}_{-0.1}$  \\
            \rule[0.7mm]{0mm}{3.5mm}
            Power-law plus absorption\tablenotemark{a}  & 199.34/240 & 1.9  & $2.8^{+0.7}_{-0.6}$ & 4.9$^{+0.3}_{-0.3}$ & 11.1$^{+0.7}_{-0.7}$ & 4.8$^{+0.3}_{-0.3}$ \\   
            \rule[0.7mm]{0mm}{3.5mm}
            Power-law plus absorption\tablenotemark{b}  & 197.99/240 & 2.13  & $3.9^{+0.7}_{-0.7}$ & 4.9$^{+0.3}_{-0.3}$ & 9.9$^{+0.6}_{-0.6}$ & 6.8$^{+0.4}_{-0.4}$ \\   
            \rule[0.7mm]{0mm}{3.5mm}
            Power-law plus absorption\tablenotemark{c} & 197.91/239 & 2.21$^{+0.30}_{-0.26}$  & $4.3^{+1.8}_{-1.5}$ & 4.9$^{+0.3}_{-0.3}$ & 9.5$^{+0.6}_{-0.6}$ & 7.7$^{+0.5}_{-0.5}$ \\[3pt]  
            \hline 
           \label{tab:fitting}
          \end{tabular}
        \begin{tablenotes}
                \vspace{5pt}
		\item Columns: (1) Model used for the spectral fitting. (2) C-statistic / degrees of freedom. (3) Best-fit or fixed photon index. (4) Intrinsic absorption, in units of $10^{23}$ cm$^{-2}$. (5) and (6) Fluxes in the observed 0.5 -- 2 and 2 -- 10 keV bands in units of 10$^{-15}$ erg cm$^{-2}$ s$^{-1}$, corrected for Galactic absorption only. (7) Rest-frame $2-10$ keV band luminosity in units of 10$^{45}$ erg s$^{-1}$, derived from the model and corrected for the Galactic absorption and intrinsic absorption (if applied), without correction for lensing magnification. 
		\item a. Spectral fitting using absorbed power-law model with a fixed $\Gamma$ of 1.9, the mean of $z \gtrsim 6$ quasars in \cite{nanni17}.
		\item b. Spectral fitting using absorbed power-law model with a fixed $\Gamma$ of 2.13, the average photon index for $z> 6$ quasars in \cite{vito19a}.
		\item c. Spectral fitting using absorbed power-law model with free $\Gamma$.
		\vspace{5pt}
	 \end{tablenotes}
\end{table*}

As discussed above, we obtain a flat spectral slope and faint X-ray luminosity for J0439+1634. To further investigate the nature of its weak X-ray emission, we perform additional spectral fits with models including more components. 
We apply spectral fitting using an absorbed power-law model ({\tt phabs*zphabs*zpowerlw} in XSPEC) to estimate the quasar intrinsic absorption, accounting for the same fixed Galactic absorption as above. 
We first assume an intrinsic photon index of $\Gamma$ = 1.9, which is the average value derived from a $z \sim 6$ quasar sample \citep{nanni17} and also consistent with that of quasars at lower redshift \citep{vignali05,just07}.
We find an intrinsic absorption of $N_{\rm H}$ of 2.8$^{+0.7}_{-0.6}$ $\times 10^{23}$ cm$^{-2}$ with a C-statistic of 199.34 for 240 degrees of freedom, resulting in an intrinsic rest-frame 2--10 keV band luminosity of $9.4^{+0.6}_{-0.6} \times 10^{43}$ $\rm erg\,s^{-1}$ (corrected for magnification). The Galactic absorption corrected flux in the 0.5--2 keV band is $4.9^{+0.3}_{-0.3} \times 10^{-15}$ $\rm erg\,s^{-1} cm^{-2}$, and the flux in the 2--10 keV band is $11.1^{+0.7}_{-0.7} \times 10^{-15}$ $\rm erg\,s^{-1} cm^{-2}$. The best-fit is shown in Figure \ref{fig:fitting}. 
As a test, we also fix $\Gamma$ to 2.13, which is the average photon index from sources with $> 30$ counts in the $z> 6$ quasar sample in \cite{vito19a}. This value is also similar to the average slope in \cite{vito19a} for sources with $ < 30$ counts and to the average value for $z> 6.5$ quasars in \cite{wang21}. Both of these two works suggest an increased photon index for quasars at $z >6$. With this $\Gamma$ value, we obtain a $N_{\rm H}$ of 3.9$^{+0.7}_{-0.7}$ $\times 10^{23}$ cm$^{-2}$ (C-stat/d.o.f. = 197.99/240), consistent with the best-fit results using $\Gamma = 1.9$ within $1\sigma$ uncertainty. 

We then perform spectral fitting using the same absorbed power-law model but with a free $\Gamma$. In this case, we find a $\Gamma$ of 2.21$^{+0.30}_{-0.26}$, with relatively large uncertainty. The best-fit intrinsic absorption, $N_{\rm H}$ = $4.3^{+1.8}_{-1.5}$ $\times 10^{23}$ cm$^{-2}$, is also more loosely constrained than in the $\Gamma$-fixed cases. We then obtain an intrinsic rest-frame 2--10 keV band luminosity of $15.1^{+1.0}_{-0.9} \times 10^{43}$ $\rm erg\,s^{-1}$. 
All fitting results show a high intrinsic column density, $N\rm_{ H} > 2\times 10^{23} cm^{-2}$, which suggests that J0439+1634 is a highly obscured quasar but not yet Compton-thick. This is the highest redshift obscured quasar with direct measurement of the absorbing column density. These fitting results are summarized in Table \ref{tab:fitting}.
If we re-compute the rest-frame 2 keV luminosity taking into account both Galactic absorption and intrinsic absorption based on the best-fits of absorbed power-law models, we find that the intrinsic 2 keV luminosity increases by $\sim$ 3 times for the fixed $\Gamma=1.9$ fit and $\sim$ 7 times for the $\Gamma$-free fit.

Since J0439+1634 is a gravitationally lensed quasar, we also consider the foreground lensing galaxy as a possible source responsible for the absorption. We use the same absorbed power-law model and fix the redshift of {\tt zphabs} component to 0.67, the best-fit redshift for the foreground lensing galaxy \citep{fan19}.
In this case, we find a column density of $N_{\rm H}$ = 5.8$^{+1.5}_{-1.3}$ $\times 10^{21}$ cm$^{-2}$ with a fixed $\Gamma = 1.9$, and $N_{\rm H}$ = 6.2$^{+2.5}_{-2.1}$ $\times 10^{21}$ cm$^{-2}$ with $\Gamma$-free fitting. The lensing galaxy absorption can reduce the quasar intrinsic X-ray luminosity by a factor of $\sim 3$. Gaseous systems with such high column density will result in damped Ly$\alpha$ (DLA) absorptions in the quasar spectrum. Due to the low redshift of the lensing galaxy and the high redshift of the quasar, the absorption features, if they exist, would be at wavelengths far bluer than the quasar Lyman break, preventing direct detection of the DLA in the quasar spectrum. However, if a DLA system with this $N_{\rm H}$ exists, we expect to see significant change in the quasar UV colors due to dust reddening: $\sim$ 1 mag extinction in the optical $z$ band, assuming the mean dust-to-gas ratio of $z \sim 0.7$ \mgii\ absorbers \citep{menard09}, which  is a tracer of low-redshift DLAs. However, the rest-frame UV spectrum of J0439+1634, after correcting for the Galactic extinction, has a slope of $\alpha_{\lambda}=-1.41$, bluer than the mean (--1.2) of $z \gtrsim 6.5$ quasars \citep{yang21}. Therefore, it is not likely that there is a high column density absorbing system from the foreground lensing galaxy. 

To consider the possible presence of an iron emission line at 6.4 keV, we also add to the absorbed power-law model a Gaussian line component. We fix the line rest-frame energy to 6.4 keV and line width to 10 eV \citep[e.g.,][]{nanni18}. All the parameters of the other components are fixed to the best-fit values from the $\Gamma$-fixed ($\Gamma$=1.9) or $\Gamma$-free fitting described above. We obtain an upper limit on line equivalent width of $EW \le 142$ eV ($\Gamma$-fixed) and $EW \le 121$ eV ($\Gamma$-free).

\subsection{An X-ray Weak BAL Quasar}
BAL quasars are known as a subclass of quasars seen with broad absorption features blueward of broad emission lines, which are thought to be evidence of  high-velocity outflows from the accretion disk.
X-ray observations of low-redshift BAL quasars show that they appear to be highly attenuated in the soft X-rays as the outflowing material could produce strong continuous absorption in X-rays. 
Previous studies of low-redshift BAL quasar samples report lower 2 keV luminosity \citep[e.g.,][]{brandt00, gibson09, luo13, luo14, liu18} than the expectation from the $\alpha_{\rm ox}$--$L_{\rm 2500 \AA}$ relation. 
Some works suggest that X-ray absorption is the primary cause of soft X-ray weakness in BAL quasars \citep[e.g.,][]{brandt00, gallagher02}, while there are also BAL quasars found as intrinsically X-ray weak quasars, for which the weakness is not entirely caused by absorption \citep[e.g.,][]{gibson08, luo14, liu18}. 

The presence of BAL features in the UV/optical spectra of J0439+1634 \citep{yang21} and its X-ray weakness relative to the $\alpha_{\rm ox}$--$L_{\rm 2500 \AA}$ relation raise questions as to whether its X-ray weakness is related to its BAL properties: is it entirely due to absorption or is the quasar also intrinsically X-ray weak?
As shown in Figure \ref{fig:aox}, J0439+1634 has X-ray properties different from the majority of high-redshift and lower-redshift quasars, but it has comparable X-ray weakness at 2 keV to that of BAL quasars. 
The drop at $< 0.8$ keV ($\sim 6$ keV in the rest-frame) shown in Figure \ref{fig:fitting} and the flat $\Gamma$ from our power-law fitting indicate absorption in the rest-frame soft band. The spectral fitting including absorption components shows a high column density, 
which suggests an absorption-caused weakness in the soft X-ray.  
As shown in Figure \ref{fig:aox}, correcting the X-ray spectra for intrinsic absorption based on the best-fits of the absorbed power-law model will only increase the 2 keV luminosity by a factor of 3 and thus yield an $\alpha_{\rm ox}$ of --1.88 if assuming $\Gamma = 1.9$. If using the $\Gamma$-free fit, although with larger uncertainties, the correction can result in an $\alpha_{\rm ox}$ of --1.76, close to the expected value ($\alpha_{\rm ox}$=--1.58) from the existing $\alpha_{\rm ox}$--$L_{\rm 2500 \AA}$ relation. Therefore, if J0439 intrinsically has a steep $\Gamma$, the absorption is a sufficient explanation for its X-ray weakness. Otherwise, there is a possibility that the intrinsic absorption is not entirely responsible for the X-ray weakness, and J0439+1634 is a candidate of intrinsically X-ray weak BAL quasar.

Intrinsically X-ray weak quasars have been suggested to be rare. \cite{gibson08} find that the fraction of sources that are under luminous by a factor of 10 is $\lesssim$ 2\% among optically selected SDSS DR5 quasars. However, \cite{liu18} suggest a higher fraction (6\% -- 23\%) of intrinsically X-ray weak quasars among BAL population. 
There are studies suggesting that BAL winds-related mechanisms could weaken or quench coronal X-ray emission \citep[e.g.,][]{proga05,luo13}. 
A very high accretion rate is also discussed as one possible reason for the intrinsic X-ray weakness of BAL quasars \citep[e.g.,][]{leighly07,luo14}. \cite{laurenti21} recently report a high X-ray weak fraction ($\sim$ 30\%) among high-Eddington ratio AGN.
For J0439+1634, its NIR spectral fitting shows an Eddington ratio ($L_{\rm bol}/L_{\rm Edd}$) of 0.6$\pm0.1$ ($L_{\rm bol} = (4.6\pm0.1)\times 10^{46}$ erg s$^{-1}$, $M_{\rm BH} = (6.3\pm0.2)\times 10^{8}\,M_{\odot}$), which is not high compared to the median value (i.e., 0.85; a mean of 1.08) for $z \sim 6.5$ quasars \citep{yang21}.
Another possible interpretation is that weak X-ray emission would not significantly ionize winds and thus intrinsically X-ray weak quasars have larger covering factors and are preferentially observed with BAL features \citep{liu18}.
On the other hand, recent work by \cite{nardini19} find a large fraction ($\sim$ 25\%) of X-ray weak quasars in a sample of luminous blue radio-quiet, non-BAL quasars at $z \sim 3$, with no clear evidence of absorption.

Previous observations also show that low-ionization BAL quasars (LoBALs), which have BAL feature on low-ionization lines, such as {\ion{Al}{3}} and \mgii, have lower 2 keV luminosity and higher absorbing column densities than high-ionization BAL quasars (HiBALs) \citep[e.g.,][]{green01, gallagher02}. \cite{sameer19} report different values of mean $\rm \Delta \alpha_{ox}$ for HiBALs ($\rm \Delta \alpha_{ox,mean}=-0.303$) and LoBALs ($\rm \Delta \alpha_{ox,mean}=-0.587$).
For J0439+1634, there is a weak \mgii\ absorption feature in its NIR spectrum, which is at the velocity corresponding to the strong \civ\ BAL absorption trough. Therefore it has low-ionization associated absorption, but with small velocity widths ($<$ 2000 km $\rm s^{-1}$) of the \mgii\ absorption compared to typical LoBALs. J0439+1634 has a $\rm \Delta \alpha_{ox}$ of --0.45, between the average HiBAL and LoBAL values.

\section{Summary} \label{sec:summary}
We present {\em XMM-Newton} observations of a gravitationally lensed BAL quasar J0439+1634 at $z=6.5$. With 130 ks of total observation time, the quasar is significantly detected with 358 net counts. Power-law only spectral fitting of the EPIC spectra yields a flat photon index ($\Gamma = 1.45^{+0.10}_{-0.09}$). This quasar is X-ray weak at rest-frame 2 keV relative to the expectation from the average $\alpha_{\rm ox}$--$L_{\rm 2500 \AA}$ relation. It is underluminous by a factor of 18 in X-ray, consistent with the behavior of BAL quasars observed at lower redshift. Spectral fitting using an absorbed power-law model suggests a high intrinsic column density, with best-fit values of $N_{\rm H} > 2 \times 10^{23}$ cm$^{-2}$ based on different $\Gamma$. J0439+1634 is therefore the first highly obscured quasar with X-ray spectroscopy in the reionization epoch, benefiting from its high lensing magnification. Accounting for a range of intrinsic absorption by assuming different $\Gamma$, we find that the 2 keV X-ray luminosity can increase by a factor of $3-7$. The possible remaining X-ray weakness indicates that J0439+1634 could be a candidate of intrinsically X-ray weak quasar.

\vspace{50pt}

\acknowledgments
This research is based on observations obtained with {\em XMM-Newton}, an ESA science mission with instruments and contributions directly funded by ESA member states and NASA. 
J. Yang, X. Fan, and M. Yue acknowledge support from US NSF grants AST 19-08284 and funding by NASA through a XMM Guest Observer program. 
F. Wang acknowledges support from NASA through the NASA Hubble Fellowship grant \#HF2-51448 awarded by the Space Telescope Science Institute, which is operated by the Association of Universities for Research in Astronomy, Incorporated, under NASA contract NAS5-26555.
G.L., M.C., and M.D. acknowledge financial support from the Italian Space Agency under grant ASI-INAF I/037/12/0, and n. 2017-14-H.O.
X.-B. Wu acknowledges support from the National Key R\&D Program of China (2016YFA0400703) and the National Science Foundation of China (11721303, 11890693).
We thank Bin Luo for providing data from published works. 

%

\vspace{5mm}
\facilities{XMM-Newton}


\software{ {\em XMM-Newton} Science Analysis Software (SAS) v19.1.0 \citep{gabriel04}, XSPEC \citep{arnaud96}}






\begin{thebibliography}{}
\bibitem[Ai et al.(2017)]{ai17} Ai, Y., Fabian, A.~C., Fan, X., et al.\ 2017, \mnras, 470, 1587
\bibitem[Arnaud(1996)]{arnaud96} Arnaud, K.~A.\ 1996, Astronomical Data Analysis Software and Systems V, 101, 17
\bibitem[Ba{\~n}ados et al.(2018)]{banados18} Ba{\~n}ados, E., Connor, T., Stern, D., et al.\ 2018, \apjl, 856, L25
\bibitem[Bisogni et al.(2021)]{bisogni21} Bisogni, S., Lusso, E., Civano, F., et al.\ 2021, \aap, 655, A109. doi:10.1051/0004-6361/202140852
\bibitem[Brandt et al.(2000)]{brandt00} Brandt, W.~N., Laor, A., \& Wills, B.~J.\ 2000, \apj, 528, 637
\bibitem[Broos et al.(2007)]{broos07} Broos, P.~S., Feigelson, E.~D., Townsley, L.~K., et al.\ 2007, \apjs, 169, 353
\bibitem[Cash(1979)]{cash79} Cash, W.\ 1979, \apj, 228, 939
\bibitem[Chartas et al.(2002)]{chartas02} Chartas, G., Brandt, W.~N., Gallagher, S.~C., et al.\ 2002, \apj, 579, 169
\bibitem[Connor et al.(2019)]{connor19} Connor, T., Ba{\~n}ados, E., Stern, D., et al.\ 2019, \apj, 887, 171
\bibitem[Connor et al.(2020)]{connor20} Connor, T., Ba{\~n}ados, E., Mazzucchelli, C., et al.\ 2020, \apj, 900, 189
\bibitem[Fan et al.(2019)]{fan19} Fan, X., Wang, F., Yang, J., et al.\ 2019, \apjl, 870, L11
\bibitem[Gabriel et al.(2004)]{gabriel04} Gabriel, C., Denby, M., Fyfe, D.~J., et al.\ 2004, Astronomical Data Analysis Software and Systems (ADASS) XIII, 314, 759
\bibitem[Gallagher et al.(2002)]{gallagher02} Gallagher, S.~C., Brandt, W.~N., Chartas, G., et al.\ 2002, \apj, 567, 37
\bibitem[Gallerani et al.(2017)]{gallerani17} Gallerani, S., Zappacosta, L., Orofino, M.~C., et al.\ 2017, \mnras, 467, 3590
\bibitem[Gibson et al.(2008)]{gibson08} Gibson, R.~R., Brandt, W.~N., \& Schneider, D.~P.\ 2008, \apj, 685, 773
\bibitem[Gibson et al.(2009)]{gibson09} Gibson, R.~R., Brandt, W.~N., Gallagher, S.~C., et al.\ 2009, \apj, 696, 924
\bibitem[Green et al.(2001)]{green01} Green, P.~J., Aldcroft, T.~L., Mathur, S., et al.\ 2001, \apj, 558, 109
\bibitem[Hasinger et al.(2002)]{hasinger02} Hasinger, G., Schartel, N., \& Komossa, S.\ 2002, \apjl, 573, L77
\bibitem[Just et al.(2007)]{just07} Just, D.~W., Brandt, W.~N., Shemmer, O., et al.\ 2007, \apj, 665, 1004
\bibitem[Kalberla et al.(2005)]{kalberla05} Kalberla, P.~M.~W., Burton, W.~B., Hartmann, D., et al.\ 2005, \aap, 440, 775.
\bibitem[Kollatschny et al.(2016)]{kollatschny16} Kollatschny, W., Schartel, N., Zetzl, M., et al.\ 2016, \aap, 585, A18
\bibitem[Laurenti et al.(2021)]{laurenti21} Laurenti, M., Piconcelli, E., Zappacosta, L., et al.\ 2021, arXiv:2110.06939
\bibitem[Leighly et al.(2007)]{leighly07} Leighly, K.~M., Halpern, J.~P., Jenkins, E.~B., et al.\ 2007, \apj, 663, 103
\bibitem[Liu et al.(2018)]{liu18} Liu, H., Luo, B., Brandt, W.~N., et al.\ 2018, \apj, 859, 113
\bibitem[Luo et al.(2013)]{luo13} Luo, B., Brandt, W.~N., Alexander, D.~M., et al.\ 2013, \apj, 772, 153
\bibitem[Luo et al.(2014)]{luo14} Luo, B., Brandt, W.~N., Alexander, D.~M., et al.\ 2014, \apj, 794, 70
\bibitem[Lusso \& Risaliti(2017)]{lusso17} Lusso, E. \& Risaliti, G.\ 2017, \aap, 602, A79. doi:10.1051/0004-6361/201630079
\bibitem[Martocchia et al.(2017)]{martocchia17} Martocchia, S., Piconcelli, E., Zappacosta, L., et al.\ 2017, \aap, 608, A51
\bibitem[Matsuoka et al.(2019)]{matsuoka19} Matsuoka, Y., Iwasawa, K., Onoue, M., et al.\ 2019, \apj, 883, 183
\bibitem[M{\'e}nard \& Chelouche(2009)]{menard09} M{\'e}nard, B. \& Chelouche, D.\ 2009, \mnras, 393, 808
\bibitem[Moretti et al.(2014)]{moretti14} Moretti, A., Ballo, L., Braito, V., et al.\ 2014, \aap, 563, A46
\bibitem[Nardini et al.(2019)]{nardini19} Nardini, E., Lusso, E., Risaliti, G., et al.\ 2019, \aap, 632, A109. doi:10.1051/0004-6361/201936911
\bibitem[Vignali et al.(2005)]{vignali05} Vignali, C., Brandt, W.~N., Schneider, D.~P., et al.\ 2005, \aj, 129, 2519
\bibitem[Vito et al.(2019a)]{vito19a} Vito, F., Brandt, W.~N., Bauer, F.~E., et al.\ 2019a, \aap, 630, A118
\bibitem[Vito et al.(2019b)]{vito19b} Vito, F., Brandt, W.~N., Bauer, F.~E., et al.\ 2019b, \aap, 628, L6
\bibitem[Vito et al.(2021)]{vito21} Vito, F., Brandt, W.~N., Ricci, F., et al.\ 2021, \aap, 649, A133
\bibitem[Page et al.(2005)]{page05} Page, K.~L., Reeves, J.~N., O'Brien, P.~T., et al.\ 2005, \mnras, 364, 195
\bibitem[Page et al.(2014)]{page14} Page, M.~J., Simpson, C., Mortlock, D.~J., et al.\ 2014, \mnras, 440, L91
\bibitem[Pons et al.(2020)]{pons20} Pons, E., McMahon, R.~G., Banerji, M., et al.\ 2020, \mnras, 491, 3884
\bibitem[Proga(2005)]{proga05} Proga, D.\ 2005, \apjl, 630, L9
\bibitem[Reed et al.(2019)]{reed19} Reed, S.~L., Banerji, M., Becker, G.~D., et al.\ 2019, arXiv:1901.07456 
\bibitem[Sameer et al.(2019)]{sameer19} Sameer, Brandt, W.~N., Anderson, S., et al.\ 2019, \mnras, 482, 1121
\bibitem[Scott et al.(2011)]{scott11} Scott, A.~E., Stewart, G.~C., Mateos, S., et al.\ 2011, \mnras, 417, 992
\bibitem[Shemmer et al.(2006)]{shemmer06} Shemmer, O., Brandt, W.~N., Schneider, D.~P., et al.\ 2006, \apj, 644, 86
\bibitem[Shen et al.(2019)]{shen19} Shen, Y., Wu, J., Jiang, L., et al.\ 2019, \apj, 873, 35 
\bibitem[Timlin et al.(2020)]{timlin20} Timlin, J.~D., Brandt, W.~N., Ni, Q., et al.\ 2020, \mnras, 492, 719
\bibitem[Schindler et al.(2020)]{schindler20} Schindler, J.-T., Farina, E.~P., Ba{\~n}ados, E., et al.\ 2020, \apj, 905, 51
\bibitem[Nanni et al.(2017)]{nanni17} Nanni, R., Vignali, C., Gilli, R., et al.\ 2017, \aap, 603, A128.
\bibitem[Nanni et al.(2018)]{nanni18} Nanni, R., Gilli, R., Vignali, C., et al.\ 2018, \aap, 614, A121
\bibitem[Venemans et al.(2015)]{venemans15} Venemans, B.~P., Ba{\~n}ados, E., Decarli, R., et al.\ 2015, \apj, 801, L11
\bibitem[Wang et al.(2019)]{wang19} Wang, F., Yang, J., Fan, X., et al.\ 2019, \apj, 884, 30
\bibitem[Wang et al.(2021)]{wang21} Wang, F., Fan, X., Yang, J., et al.\ 2021, \apj, 908, 53.
\bibitem[Weisskopf et al.(2007)]{weisskopf07} Weisskopf, M.~C., Wu, K., Trimble, V., et al.\ 2007, \apj, 657, 1026
\bibitem[Yang et al.(2019a)]{yang19a} Yang, J., Venemans, B., Wang, F., et al.\ 2019a, \apj, 880, 153
\bibitem[Yang et al.(2019b)]{yang19b} Yang, J., Wang, F., Fan, X., et al.\ 2019b, \aj, 157, 236
\bibitem[Yang et al.(2021)]{yang21} Yang, J., Wang, F., Fan, X., et al.\ 2021, arXiv:2109.13942
\bibitem[Yue et al.(2021)]{yue21} Yue, M., Yang, J., Fan, X., et al.\ 2021, \apj, 917, 99
\end{thebibliography}



\end{document}